\documentclass{PoS}

\PoS{PoS(LAT2005)140}

\usepackage{graphicx}
\usepackage{amsmath}

\newcommand{\bee}{\begin{equation}}
\newcommand{\ee}{\end{equation}}
\newcommand{\beea}{\begin{eqnarray}}
\newcommand{\eea}{\end{eqnarray}}
\newcommand{\gfive}{\gamma_5}
\newcommand{\sign}{{\rm sign}}

\def\Tr{{\rm Tr}}

\title{Dynamical overlap fermions:
 simulation and physics results}

\ShortTitle{Dynamical overlap fermions: techniques and results }

\author{\speaker{Stefan Schaefer}\\
        University of Colorado\\
        E-mail: \email{schaefer@pizero.colorado.edu}}

\author{Thomas DeGrand\\
        University of Colorado\\
        E-mail: \email{degrand@pizero.colorado.edu}}

\abstract{
We summarize our recent investigations of lattice QCD with dynamical overlap fermions.
We sketch algorithmic issues and our approach to solving them. We show our measurement of the topological
susceptibility.
We describe 
a computation of the chiral condensate using an analysis of the distribution of
eigenmodes of the Dirac operator and Random Matrix Theory.
}

\FullConference{XXIIIrd International Symposium on Lattice Field Theory\\
                 25-30 July 2005\\
                 Trinity College, Dublin, Ireland}

\begin{document}

\section{Introduction.}

More than twenty years' experience of the lattice gauge community has taught us
that it is always a good thing to have a bare action which respects symmetries,
because then  no fine tunings are required to preserve the symmetries at distances
much greater than the cutoff scale. For this reason, essentially all lattice simulations of
gauge theories perform simulations with gauge invariant lattice actions,
and there is never any discussion about trading a small violation of gauge invariance in the simulation
for larger volume or an apparently more efficient simulation algorithm.
People have, however, always been willing to sacrifice chiral symmetry in their choice of lattice
discretization.
This seems somehow asymmetric:
Why not perform simulations with lattice actions which
preserve exact  $SU(N_f)\times SU(N_f)$ chiral symmetry?

The advantages of this approach are obvious: One does not have to
 separate the physical explicit
chiral symmetry breaking from a nonzero quark mass from the unphysical
chiral symmetry breaking induced by lattice artifacts.
The flavor content of the theory being
simulated is unambiguous. The index theorem is theoretically clean.
The topological charge can be measured to be exactly what the dynamical fermions
 see during the simulation,
not something which is determined  by some post-processing procedure.  And because
the action preserves symmetries, correlation functions obey Ward identities which
considerably simplify their theoretical analysis. For example,
one does not have to spend any time measuring
(and trying to remove) lattice-artifact additive mass renormalization or operator mixing.

The way to do this is well known: use a lattice action which encodes
 the Ginsparg-Wilson\cite{Ginsparg:1981bj} relation,
an overlap\cite{Neuberger:1997fp,Neuberger:1998my}
action. This article is a summary of our experiences with simulations
of two flavors of dynamical overlap fermions, using a version of the algorithm
 of Fodor, et al\cite{Fodor:2003bh}.
It is a condensation of our two recent papers, Refs. \cite{DeGrand:2004nq,DeGrand:2005vb},
plus a little newer material.

\section{Algorithmic Issues}

Simulations with dynamical overlap have (at least) two problems:
\begin{itemize}
\item
 They run so slowly
\item
Changing topology is hard
\end{itemize}

Our method of attack for the first problem is to replace the
usual  link variable gauge connection by a fat link. It has been known
since at least 1998 \cite{DeGrand:1998pr}
 that fat links improve the chiral properties of
non chiral fermion actions (and the flavor symmetry properties
of staggered fermions). The bottleneck has always been to find a
 smearing method
which can be used in a molecular dynamics update, where the evaluation
of the force requires a fat link which is differentiable with respect to its
component thin links. Formulations like the Asqtad link solve this problem
by ``following the paths,'' but this does not give as much improvement
as one would like. Our solution was provided by  
Peardon and Morningstar\cite{Morningstar:2003gk}
with the ``stout link'' (invented in a Dublin public house): a multilevel
blocking which is fully differentiable. In our runs, it pushes the thin link
plaquette $\Tr U_p \sim 1.7$ up to about 2.8 (with two levels of smearing with $\rho=0.15$).
The number of Dirac operator matrix times vector multiplies per trajectory is reduced by
about an order of magnitude compared to simulations with a thin link gauge connection.
The physics reason for this speedup is that fattening reduces the number of small eigenmodes
of the kernel operator, improving its  conditioning number.

With a stout link, the fermions decouple from the UV fluctuations of the gauge field, and 
the mean size of the fermion force is reduced to about an order of magnitude smaller than 
the gauge force. This makes a multiple time scale integration algorithm very attractive.
We run with the Sexton-Weingarten\cite{Sexton:1992nu}
 form of this updating, taking the integration
time step for the gauge fields to be 1/12 of that for the fermions.

The square of the Hermitian overlap operator projected on one chiral
sector is given by
\bee
H^2_\sigma(m)= 2 (R_0^2-\frac{m^2}{4})
P_\sigma\left[1+\sigma \sum_i \epsilon(\lambda_i)
 |\lambda_i\rangle\langle \lambda_i| \right]P_\sigma+m^2
\label{eq:shift}
\ee
with $R_0$ the radius of the Ginsparg--Wilson circle,
 $P_\sigma=\frac{1}{2}(1+\sigma \gfive)$ the projector
on chirality $\sigma$ and $h(-R_0)$ the Hermitian kernel operator.
The sign function $\epsilon(h(-R_0))$ is
here given in its spectral representation.

Because of the sign function in its definition, the effective action of the overlap
operator has a discontinuity. It occurs when  one 
eigenvalue of the kernel $h(-R_0)$ changes sign during the 
molecular dynamics evolution. These are the surfaces
in the space of the gauge fields
on which the topology as defined by the index theorem changes by one unit.
Ref.~\cite{Fodor:2003bh} gives a prescription of how
to account for this discontinuity
in the HMC algorithm. One essentially measures the height $\Delta S$
of the step in the action 
(the potential of our Hamiltonian equations of motion)
and if the momentum perpendicular to the surface is large enough one reduces it
as one would do in classical mechanics. We will call this a ``refraction'' in 
the following. If the perpendicular momentum is too small, we flip it, and thus
reflect the trajectory. With $N$ the vector normal to the surface
momenta $\pi$ are thus updated by
\bee
\Delta \pi =
\begin{cases}
-N \; \langle N |\pi \rangle + N \; \sign  \langle N | \pi \rangle \; 
 \sqrt {\langle N | \pi \rangle^2-2 \Delta S_f}
& \text{if  $\langle N | \pi \rangle^2>2 \Delta S_f $}\\
-2  N \langle N | \pi \rangle &  \text{if $\langle N | \pi \rangle^2\leq 2 \Delta S_f$}
\end{cases}
\label{eq:ref}
\ee

The discontinuity $\Delta S$  of the effective action is caused by one eigenvalue 
changing sign, thus making the replacement
\bee
H^2_\sigma(m) \longrightarrow H_\sigma^2\pm (4R_0^2-m^2) P_\sigma|\lambda_0\rangle\langle \lambda_0|P_\sigma
\ee
with $|\lambda_0\rangle$ the zero mode. The corresponding step in the effective action
can be evaluated using the Sherman--Morrison formula~\cite{Golub}
\bee
\Delta \left [ \langle \phi| P_\sigma \frac{1}{H_\sigma(m)^2}P_\sigma| \phi \rangle \right] = 
\mp 
\frac{(4R_0^2-m^2)}
{1\pm (4R_0^2-m^2) \langle \lambda_0|P_\sigma H^{-2}_\sigma(m) P_\sigma|\lambda_0 \rangle}
|\langle  \phi |P_\sigma\frac{1}{H_\sigma(m)^2}P_\sigma| \lambda_0\rangle|^2 \ .
\label{eq:sherman}
\ee
Interestingly, for the overlap  not only can one  compute the step in the effective action, but one can
also give a closed form expression
 for the change in the fermionic determinant due to the change in 
topology:
\bee
  \frac{\det \tilde H^2_\sigma(m)}{\det  H^2_\sigma(m)}
 =  1\pm(4R_0^2-m^2) \langle \lambda_0 |P_\sigma \frac{1}{H^2_\sigma(m)} P_\sigma | \lambda_0\rangle \ .
 \label{eq:stepdet}
\ee

In the actual simulation one faces the problem that the trajectories reflect most of the
time off the zero eigenvalue surface and one never changes topology.
The reason is that the change in the determinant from the starting configuration
to the 'current' configuration in the MD evolution is only approximated well
as long as the Dirac operator is similar to the starting one.
The fluctuations are small as long as one has not changed topology.
However, this is definitely not
the case for the operator in a different topological sector. Since $\exp(-\phi^+1/H^2\phi)$
averages to the change in the determinant from the starting configuration to the end, 
large fluctuations mean that most of the time $\exp(-\phi^+1/H^2\phi)\approx 0$ and
the effective action is thus large, whereas only a few times do we get small effective actions.
 These 
two observations combine to give large $\Delta S$ most of the time, and thus
a large auto-correlation time in the topological charge.

To reduce fluctuations, we used the method proposed in
Refs.~\cite{Hasenbusch:2001ne,Hasenbusch:2002ai},
which consists of rewriting the fermion determinant as
as
\bee
\det H^2(m) =  \det H^2(m_{N_p})\prod_{i=1}^{N_p-1} \det \frac{H^2(m_i)}{H^2(m_{i+1})}
\label{eq:det}
\ee
with $m_1=m$ and $m_i<m_{i+1}$ with suitably chosen larger masses.
 In this method, only determinant ratios are evaluated
using pseudo-fermions for the light quark masses. The change in the spectra
while changing topological sector of the ratio $H(m)/H(m')$ is expected to be
less dramatic than the change of the spectrum of  $H(m)$. Only the
determinant of $H(m_N)$  is evaluated directly.
However, for a large mass $m_N$ the spectrum of $H^2$ is confined to a smaller
region between $m_N^2$ and  $4 R_0^2$
and the change in the spectrum therefore less drastic than for a smaller mass.
One or two extra pseudo-fermion fields ($N_p=2,3$ in Eq. \ref{eq:det}) help some, but do not
solve the problem.

To quantify our difficulty, we compare in Fig.~\ref{fig:1} the discontinuity of the effective
action with the physical step from the fermion determinant. We subtracted
the relevant quantity from the normal component of the momentum so  that
positive values correspond to reflections whereas the topology changes for negative values.
We observe that the physical discontinuity would allow for significantly more changes in
topology than the step in the effective action does.

\FIGURE[tb]{
\includegraphics[width=0.3\textwidth, angle=-90, clip]{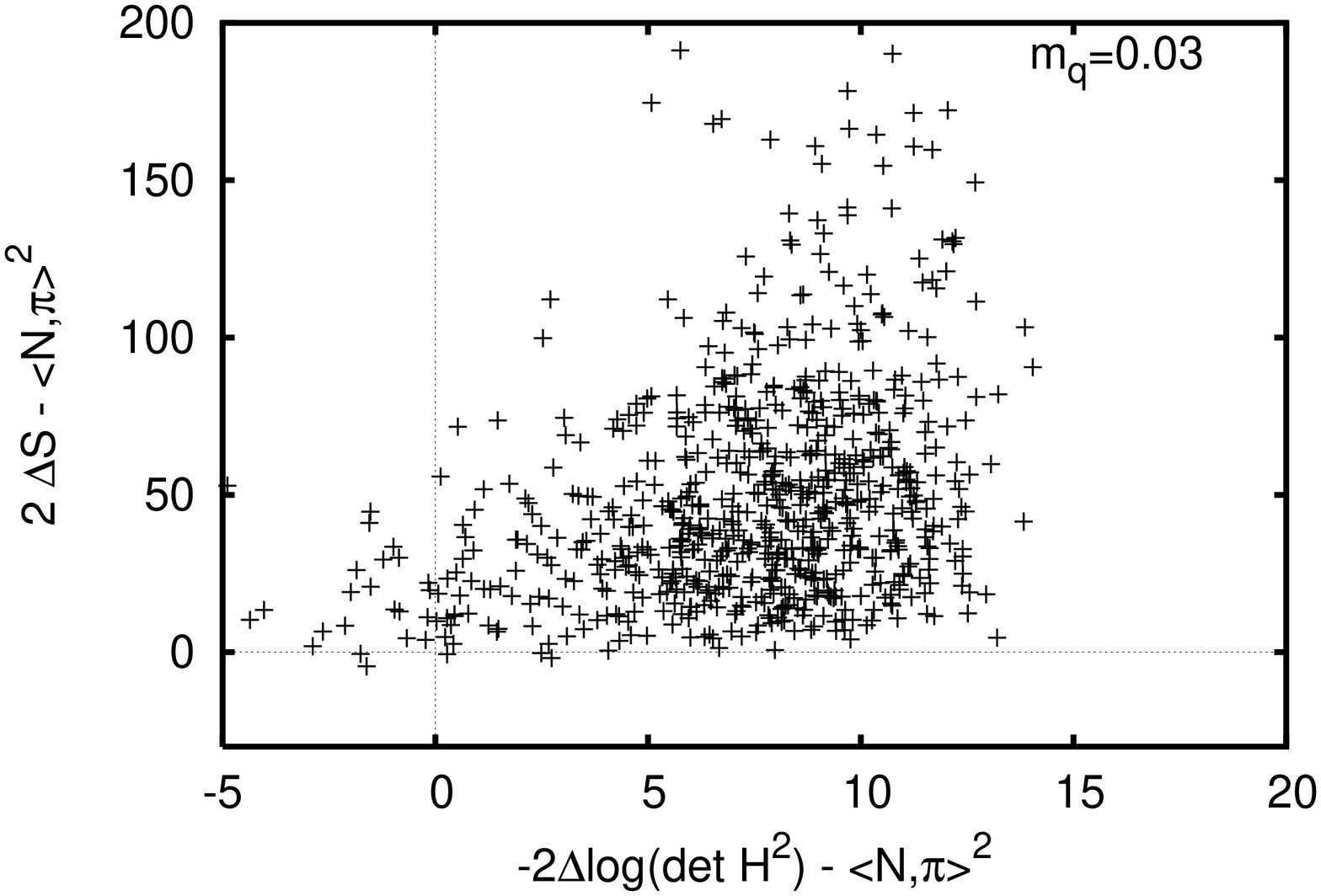}
\includegraphics[width=0.3\textwidth, angle=-90, clip]{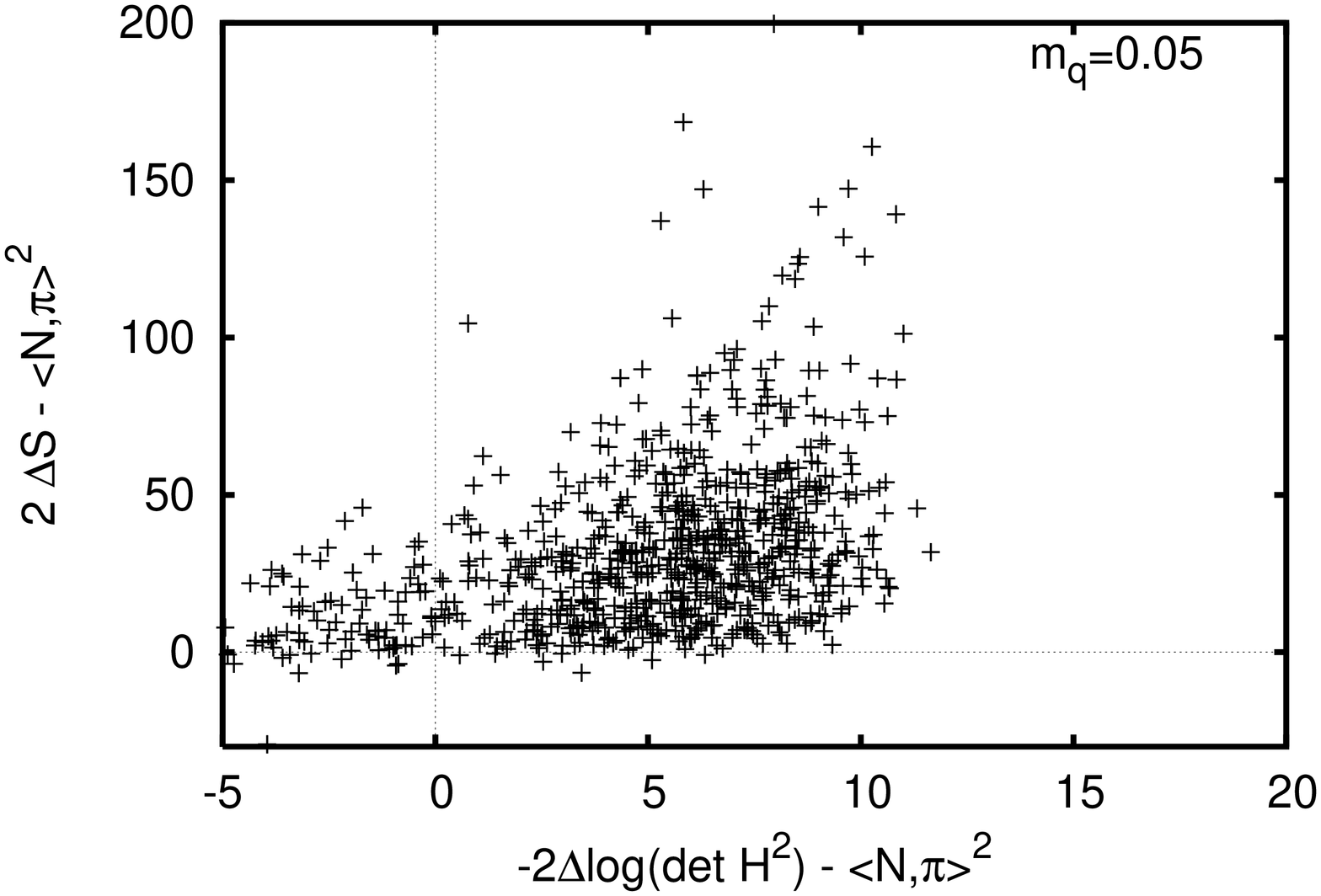}
\caption{The stochastic estimate of the height of the step compared to the actual change
in the logarithm of the determinant from a subset of our ensemble.
We subtracted the normal component of the momentum squared (which is typically less than 10) such
that negative values mean refraction and positive ones reflection. For mass $m_q=0.03$ on
the left we have a number of events in the upper left quadrant that would have tunneled with the
exact change of the determinant and only a few that actually tunneled (in the two lower
quadrants). For $m_q=0.05$ the picture is similar, even though there are more tunneling events.
\label{fig:1}
}
}

\begin{figure}
\begin{center}
\includegraphics[width=0.4\textwidth,clip]{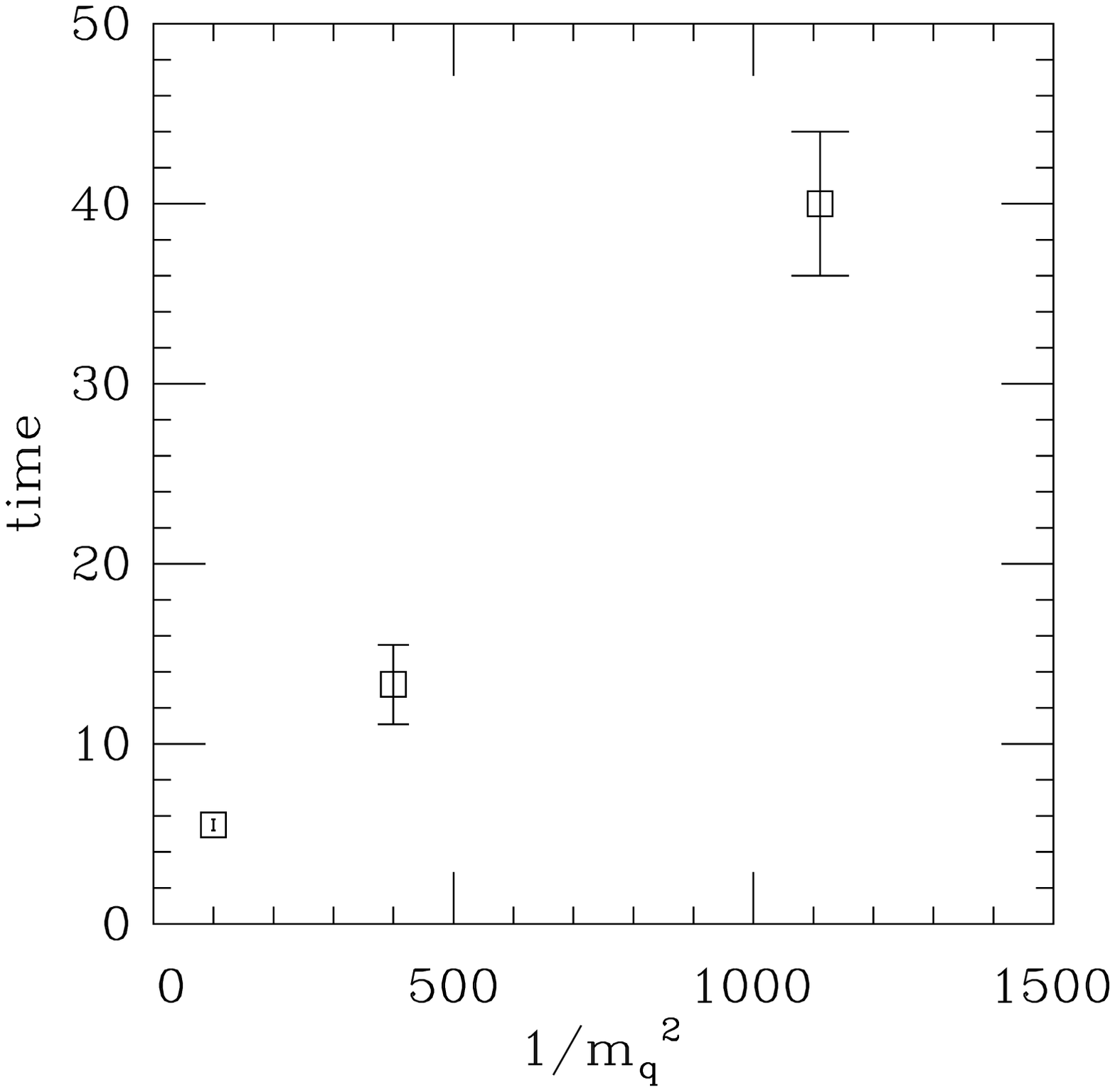}
\end{center}
\caption{Monte Carlo simulation time between topology changes versus quark mass.
\label{fig:changevmq}
}
\end{figure}

The low correlation between the estimator and the physical step height Eq.~(\ref{eq:stepdet})
shows  up in the large auto-correlation time of the topological charge. Even
though part of it is physics --- lighter quarks make it harder to get from
 $\nu=0$ to $\nu=\pm 1$ --- the height of the step grows with $1/m^2$ instead
of the expected determinant ratio, $\log~m$. Since the normal component of the momentum
is roughly independent of the quark mass, it becomes more and more difficult to change topology.
Indeed, Fig.~\ref{fig:changevmq} shows that the mean time between
topological changes varies inversely with the square of the quark mass.
The large auto-correlation time for the topology is
a phenomenon  that is also known with other fermions, e.g. improved  staggered quarks.
To the extent that these formulations know about topology, the step in the fermion action
for the overlap might be replaced for them by a steep region which approximates the step.
The result is the same: if the approximation of the determinant is bad, the step is
overestimated most of the time and one does not change topology.

\section{The Topological Susceptibility}
In our second paper we made  rough calculations of the topological
susceptibility and
chiral condensate using
eigenmodes of the Dirac operator. We made simulations on $8^4$ lattices, at a lattice
spacing of about $a\sim 0.16$ fm, with three quark masses, $am_q=0.03$, $0.05$, and $0.1$.
These lattices are really too small for physics, but they illustrate
the useful features of a calculation with overlap fermions.

We determined the string tension from the heavy quark potential and the Sommer parameter,
from Wilson loops of temporal extent $t=2$ and 3. The two measurements are not consistent,
but we performed tests on $8^4$ and $12^4$ quenched lattices which showed that the $t=3$
potentials were consistent with ones from further separations. So we used their fit values.
The lattice spacing varies by about ten percent as we change the quark mass.

\begin{figure}
\begin{center}
\includegraphics[width=0.4\textwidth,clip]{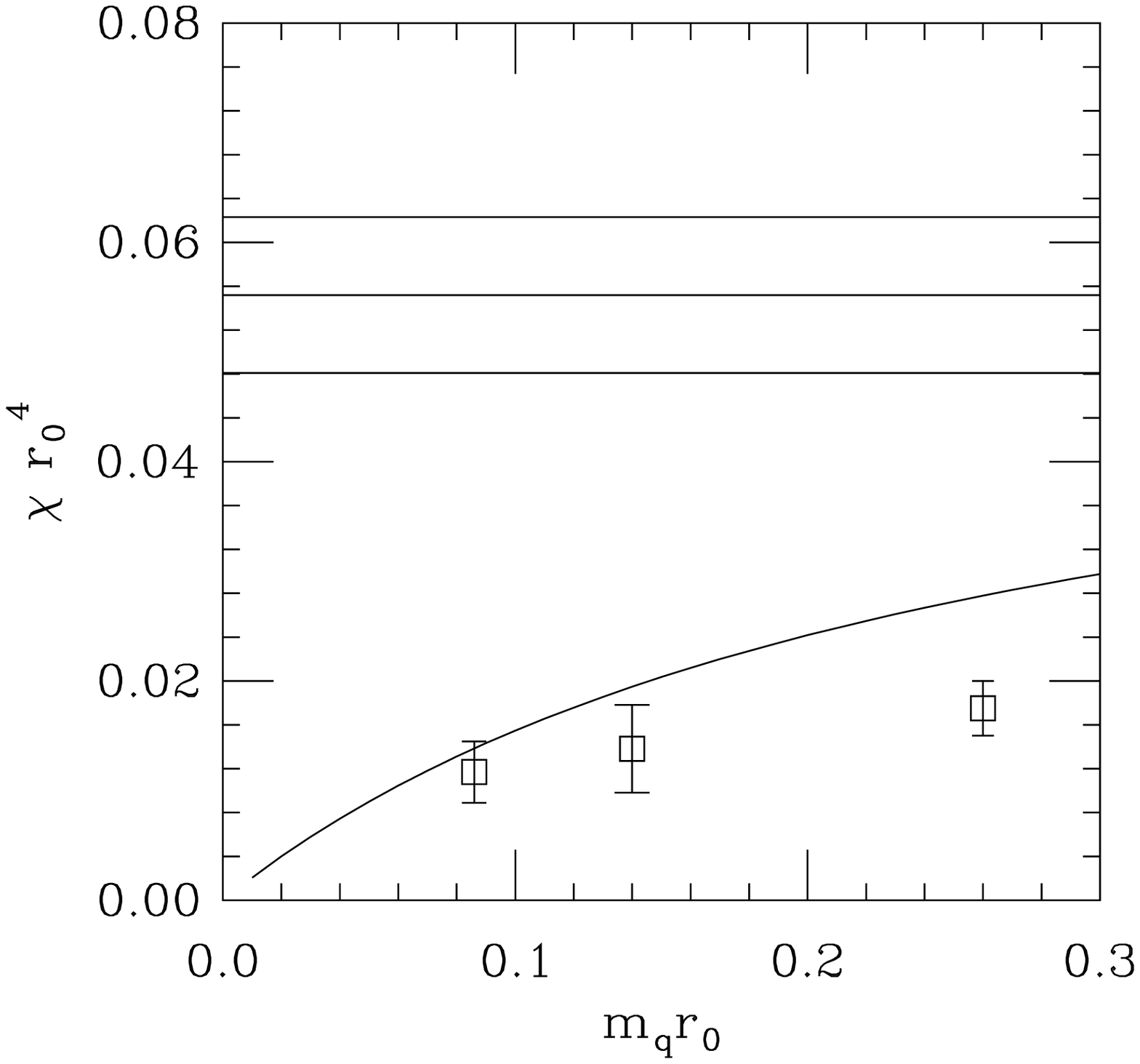}
\end{center}
\caption{Topological susceptibility versus quark mass, in units of
$r_0$.
The curved line is the D\"urr interpolating formula, Eq. \protect\ref{eq:durr}.
The three horizontal lines give the quenched value and its error.
}
\label{fig:topomr0}
\end{figure}

One picture, Fig.~\ref{fig:topomr0},
 illustrates our measurement of the topological susceptibility $\chi$.
 We take our measurements of $r_0/a$ and the topological charge time history
 to compute $\chi r_0^4$. We have computed the
lattice-to-$\overline{MS}$ matching factor in perturbation theory and use it to
convert the quark masses to their $\mu=2$ GeV
$\overline {MS}$ values.
D\"urr\cite{Durr:2001ty} has presented a phenomenological interpolating formula
for the mass dependence of the topological susceptibility,
in terms of the condensate $\Sigma$ and quenched topological susceptibility $\chi_q$,
\bee
\frac{1}{\chi}=  \frac{N_f}{m_q\Sigma}+ \frac{1}{\chi_q}.
\label{eq:durr}
\ee
Taking $\Sigma$ from our RMT analysis in the next section ( ${r_0}^3 \Sigma\sim0.43$)
produces the curve shown in the figure.

Most published measurements of the topological susceptibility present them as a function of the pseudoscalar
mass. Since we don't have spectroscopy, we can't do that. We can, however, use the D\"urr
curve as a fiducial, since most published measurements of the topological susceptibility
present it, too. Our data (as well as that of Ref. \cite{Fodor:2004wx}) lies below the D\"urr
curve.
Most measurements with nonchiral actions lie above it. (See, for example the figures in
Ref. \cite{Hasenfratz:2001wd} or \cite{Allton:2004qq}). Since our quenched results give
a value typical of simulations on larger lattices, $\chi \sim (190$ MeV$)^4)$, we don't think we are seeing
a finite volume effect.

\section{The Condensate from Eigenmode Distributions}

It was proposed more than a decade ago that the distribution of the low-lying eigenvalues
of the QCD Dirac operator in a finite volume can be predicted by random matrix
theory (RMT)~\cite{Shuryak:1992pi,Verbaarschot:1993pm,Verbaarschot:1994qf}.
Since then this hypothesis has received impressive support from lattice calculations,
mainly quenched simulations
\cite{Berbenni-Bitsch:1997tx,Damgaard:1998ie,Gockeler:1998jj,Edwards:1999ra,Bietenholz:2003mi,Giusti:2003gf},
but also some dynamical ones using staggered quarks
\cite{Berbenni-Bitsch:1998sy,Damgaard:2000qt}.

Typically, the predictions are made in the so-called epsilon regime,
 for which $1/\Lambda \ll L \ll 1/m_\pi$ with $\Lambda$ a typical hadronic scale.
However, it has been found that they describe the data in a wider range.
Two recent large scale studies, e.g.,  using the overlap operator on quenched
configurations~\cite{Bietenholz:2003mi,Giusti:2003gf},  needed lattices
with a length larger than $1.2-1.5~{\rm fm}$ for   RMT predictions match the result of the
simulation. Our dynamical lattices have a spatial extent of about $1.3~{\rm fm}$.
As we will see, random matrix theory describes our low-lying Dirac spectra quite well.

Our analysis is based on
the distribution of the $k$-th eigenmode from RMT as presented
in Ref.~\cite{Damgaard:2000ah} and successfully compared to simulation results
in Ref.~\cite{Damgaard:2000qt}. The prediction is for the distribution of the
dimensionless quantity $\zeta=\rho \lambda_k \Sigma V$ in each topological sector, with
$\lambda_k$ the $k$-th  eigenvalue of the Dirac operator,
$\Sigma$ the chiral condensate, and $V$ the volume
of the box. 
 The quantity $\rho$ is
the one-loop finite volume correction, $\rho = 1 + c/(f_\pi L)^2$ where $c$ is
a ``shape factor.''
These distribution are universal and depend only on
 the number of flavors, the topological charge and the dimensionless
quantity $m_q \Sigma V$.

By comparing the distribution of the eigenmodes with the RMT
prediction one can thus measure the chiral condensate $\Sigma$.
The main advantage of this method  is that it gives the
zero quark mass, infinite volume condensate directly. The validity of the approach
can be verified comparing the shape of the distribution for the various modes and topological
sectors. The main uncertainty comes from a too small volume which causes deviations in the
shape, particularly for the higher modes.

In Fig.~\ref{fig:rmt} we show the distribution of the two lowest eigenmodes of the overlap
operator (scaled by $\Sigma V$)
measured on the $\nu=0$ and $\nu=\pm1$ parts of the  $am_q=0.03$ and $am_q=0.05$ ensembles.
We fit the RMT prediction from Ref.~\cite{Damgaard:2000ah} to these distributions.
The prediction agrees overall well with the measured distribution given the low statistics.
However, the distribution of the lowest mode in the $\nu=0$ sector seems to have a tail at larger
$\lambda \Sigma V$ that does not match the prediction. This could be an effect of the small
 volume.
We also show the prediction for the distribution of the third mode from our fitted values
of $\Sigma V$ in the third column of Fig.~\ref{fig:rmt}. The RMT curve and the data, again,
agree quite well. However for the $|\nu|=1$ sector, the curve seems to be on the right of
 the data.
This is probably a sign of the breakdown of RMT for eigenvalues larger than the Thouless
energy~\cite{Osborn:1998nm,Gockeler:1998jj}.

\FIGURE[tbh]{
\includegraphics[width=0.5\textwidth,clip,angle=-90]{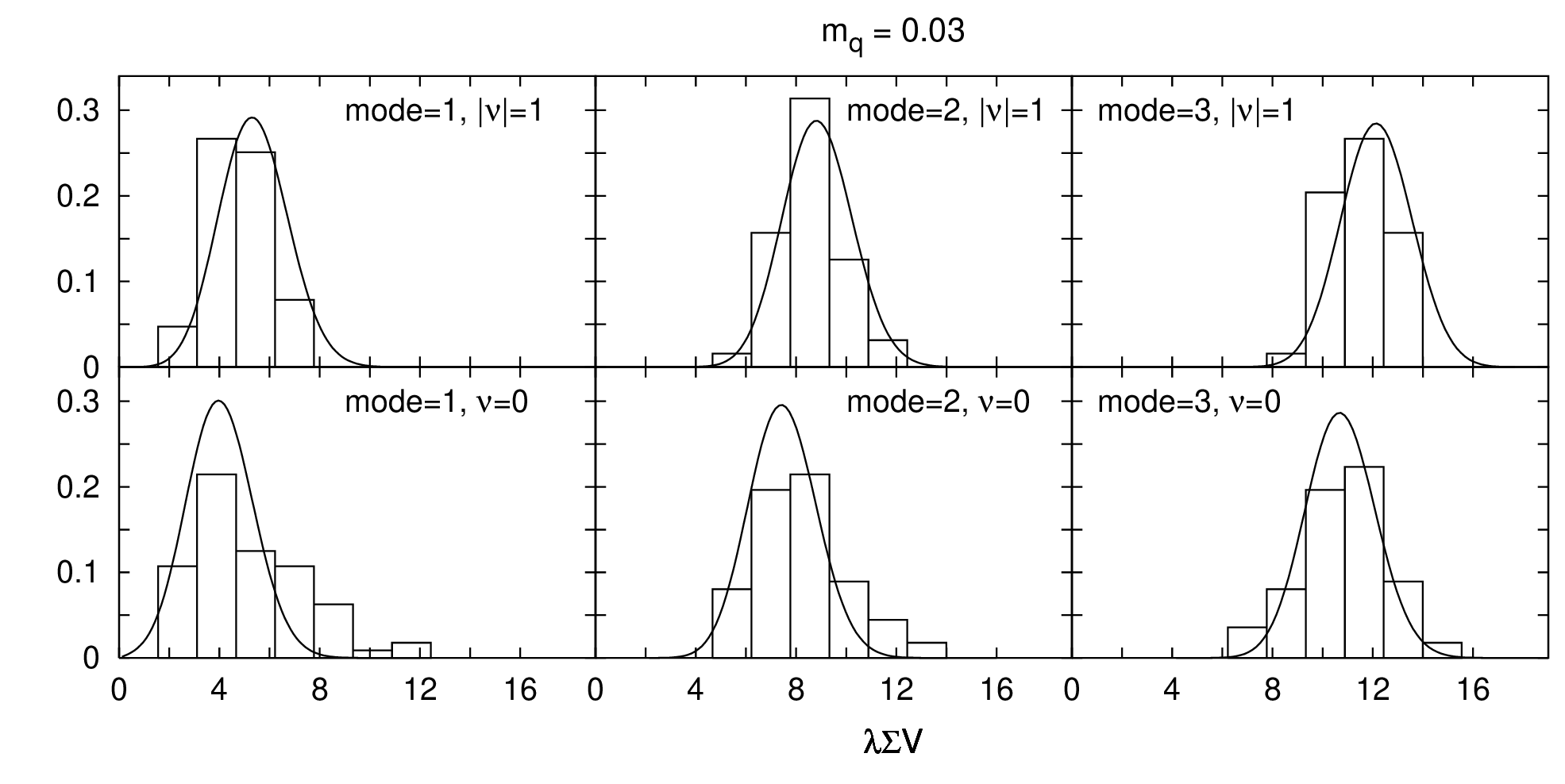}
\includegraphics[width=0.5\textwidth,clip,angle=-90]{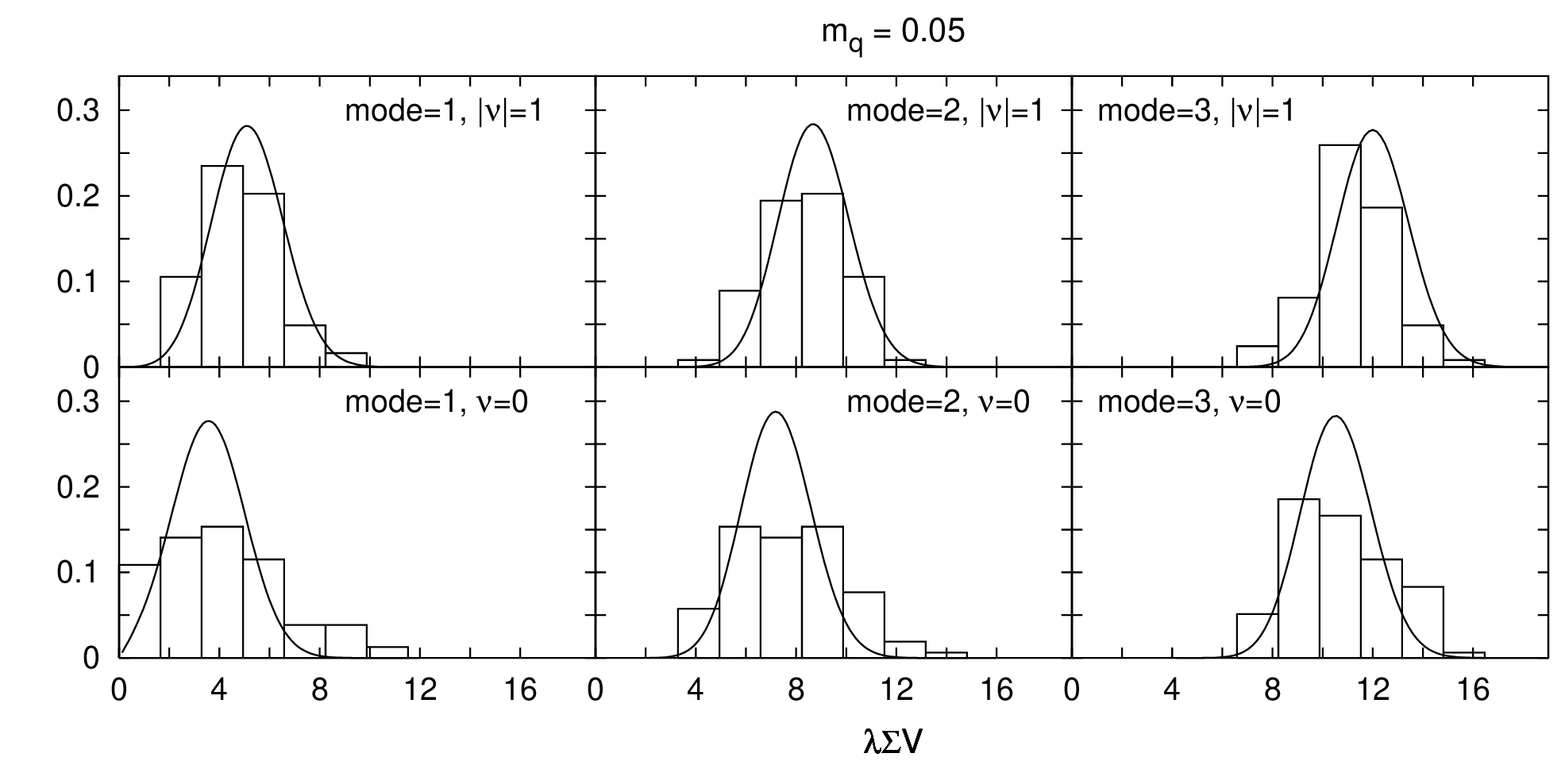}
\caption{\label{fig:rmt}Distribution of the lowest two eigenmodes of the
Dirac operator for our ensemble for
$\nu=0,\pm1$. The lines are   fits of the random matrix
theory prediction to the data for the two lowest modes.
The lines for the third mode are predictions.}
}
\TABLE[htb]{
\begin{tabular}{c|c}
$am_q$ \  &\  $\rho \Sigma r_0^3$  \\
\hline
0.05 & 0.40(2) \\
0.03 & 0.44(2) \\
0.01 & 0.38(2) \\
\hline
\end{tabular}
\caption{\label{tab:tab1} Condensate versus quark mass.}
}
A combined fit to $\nu=0,1$, $n=1,2$ at each $m_q$ (four distributions fit simultaneously)
gives the results shown in Table \ref{tab:tab1}. In our small volumes, 
and using the physical value for $f_\pi$
 (93 MeV), $\rho \sim 1.4$, which is uncomfortably large. Dividing it out boldly
gives $\Sigma \sim (280$ MeV$)^3$.

After Ref. \cite{DeGrand:2005vb} appeared, we performed some simulations at lower
quark mass, $am_q=0.01$. We restricted the topological sector to $\nu=0$ by forbidding
refractions. It is unknown whether a particular topological sector is simply
connected or not (in the latter case, forbidding topology changes might make the simulation
non-ergodic).
We ran two separate molecular dynamics streams to look for any obvious discrepancies.
We did not see any:
In the two streams, the plaquette and string tension parameters were
consistent within statistical uncertainties; nothing looks unusual. We also did some
running at $am=0.005$ (or a 5 MeV quark mass), though not with enough statistics
to fit the condensate. In both cases the code ran stably and quietly.

 Fig. \ref{fig:rmt01}
shows the distributions and fit to the condensate from this data set. 

\FIGURE[tb]{
\includegraphics[width=0.33\textwidth,clip,angle=-90]{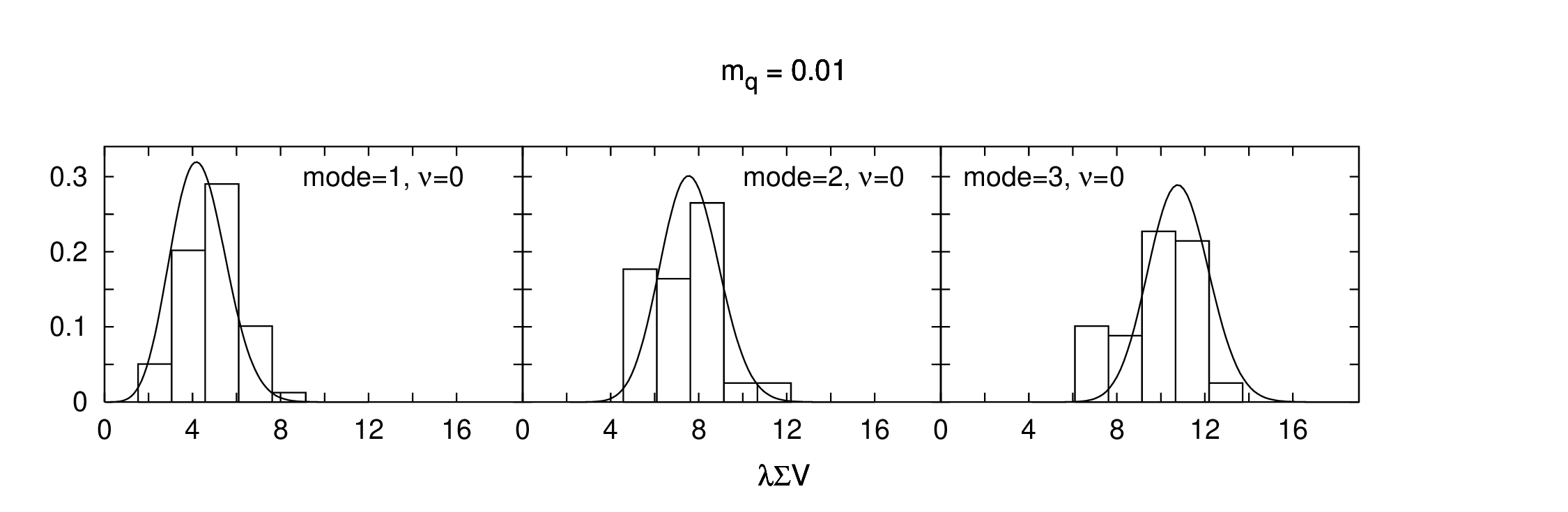}
\caption{\label{fig:rmt01}Distribution of the lowest two eigenmodes of the
Dirac operator for our ensemble for
$\nu=0$ at $am_q=0.01$. The lines are   fits of the random matrix
theory prediction to the data for the two lowest modes.
The lines for the third mode are predictions.}
}

\section{Conclusions}

Simulations with dynamical overlap fermions are poised to begin producing physics results.
We are presently simulating on larger volumes in order to make a more reliable calculation
of the condensate.
We also continue to groom our algorithms. 
We believe that there are many more tricks to be found and encourage others to work
on dynamical simulations with overlap fermions.
The physics payoffs are potentially very high.

This work was supported by the US Department of Energy.

\end{document}